\documentclass[a4,12pt]{article}
\usepackage{graphicx}
\usepackage{amsmath}
\usepackage{amssymb}
\usepackage{hyperref}
\usepackage{indentfirst}
\usepackage{makeidx}
\usepackage[latin1]{inputenc}
\title{Topological spin excitations induced by an external magnetic field coupled to a surface with rotational symmetry}
\author{Vagson L. Carvalho-Santos $^1$, Rossen Dandoloff $^2$
\\\small$^1$Instituto Federal de Educa\c c\~ao, Ci\^encia e Tecnologia Baiano - Campus Senhor do Bonfim\\\small 48970-000 Senhor do Bonfim, Bahia, Brazil\\\small$^2$Laboratoire de Physique Th\'eorique et Mod\'elisation, Universit\'e de Cergy-Pontoise\\\small
95302 Cergy-Pontoise, France}

\date{}

\begin{document}
\maketitle
\begin{abstract}
We study the Heisenberg model in an external magnetic field on curved surfaces with rotational symmetry. The Euler-Lagrange static equations, derived from the Hamiltonian lead to the inhomogeneous double sine-Gordon equation (DSG). However, if the magnetic field is coupled with the metric elements of the surface, and consequently, its curvature, the homogeneous DSG appears and a $2\pi$-soliton is obtained as a solution for this model. In order to obey the self-dual equations, surface deformations are predicted at the sector where the spins point in the opposite direction to the magnetic field. The model was used to particularize the characteristic lenght of the 2$\pi$-soliton for three specific rotationally symmetric surfaces: the cylinder, the catenoid and the hyperboloid. Fractional 2$\pi$-solitons must appear on finite surfaces, as the sphere, torus and barrels, for example.\\\\
\textbf{Keywords}: Classical spin models, solitons, Curvature, Heisenberg Hamiltonian\end{abstract}



\section{Introduction and Motivation}
In the last decades, the study of relations between geometry and the physical properties of condensed matter systems (CMS) has attracted large attention. On the one hand, the growing capacity for fabricating and manipulating nanoscale devices with different geometries, such as quasi two-dimensional exotic shapes like the M\"obius stripe \cite{Mobius-Nature}, torus \cite{Yazgan-Torus} and asymmetric nanorings \cite{Zhu-PRL-96}, brings the possibility to develop and test theoretical models in several branches of CMS, e.g., nanomagnetism, nematic liquid crystals, graphene devices and topological insulators. On the other hand, theoretical models predict the appearing of collective modes, which are strongly influenced by the curvature of the substrate. For instance, particle-like excitations, like vortices and solitons, appear in several contexts, as superconductors, ferromagnetic nanoparticles, nematic liquid crystals, Bose-Einstein condensates \cite{Chaikin-Book,Mermin-Review} and
  colisionless plasmas \cite{BJP-REF1}. It has been shown that these particle-like excitations, which have topological character, interact not only with each other, but also with the curvature of the substrate \cite{Curv-Effect}. 

In the case of ferromagnetic materials, cylindrical nanomagnets with a vortex as the magnetization groundstate have been considered as candidates to be used in logic memory, data storage, highly sensitive sensors \cite{Dai-Appl} and cancer therapy \cite{Cancer-Therapy}. Theoretical works have shown that the energy and stability of these topological excitations depend on the curvature of the ferromagnetic nanoparticle \cite{Vagson-JAP-105} and their dynamical properties are affected by both: the interaction with curve defects appearing during the fabrication of the nanomagnets \cite{Apolonio-JAP}; and the presence of magnetic/non-magnetic impurities in these magnetic nanoparticles \cite{Pablo-JAP}. 

In addition, vortices can appear like solutions to the continuous Heisenberg Model on two-dimensional systems. In Refs. \cite{torus}$-$\cite{cone}, the authors have used this model to analyse the dynamic and static properties of vortices and have shown that the energy of these excitations is closely linked to the characteristic length of the considered geometry. Besides, the vortex energy presents a divergence in simply-connected surfaces, which can be controlled by the development of an out-of-plane component in the vortex core, so called the vortex polarity. Soliton-like solutions have also been considered in the above cited works and it has been shown that their characteristic lengths depend on the length scale of the surface. For finite surfaces, fractional/half-soliton solutions have been obtained \cite{pseudosphere,Saxena-PRB66}. 

Note that CMS in the presence of an external field exhibit changes in the behavior of collective modes and in their elastic properties. For example, an external magnetic field can be used to deform magnetoelastic metamaterials, which is achieved by providing a mechanical degree of freedom so that the electromagnetic interaction in the metamaterial lattice is coupled to elastic interaction \cite{Lapine-Nature}. Furthermore, by combining the curvature effects with magnetic fields, the molecular alignment of flux lines of the nematic director of nematic liquid crystals can be reoriented or switched between two stable configurations \cite{Napoli-PRL-108}. Lastly, it has been shown that the curvature of graphene bubbles can be controlled by applying an electric field \cite{georgiou-AIP}. 

In this context, Saxena \textit{et al.} have considered the exchange and Zeeman terms in the magnetic energy calculations for cylindrical surfaces, and they have shown that the interaction of an external magnetic field with a cylindrical magnetoelastic membrane has a 2$\pi$ soliton-like solution, which induces a deformation (pinch) at the sector where the spins are pointing in the opposite direction to the magnetic field \cite{Saxena-PRB-58}. However, if a constant magnetic field is applied on an arbitrary curve surface, with rotational symmetry, the non-homogeneous double sine-Gordon equation (DSG) is obtained and numerical solutions are demanded \cite{Dandoloff-JPhysA}. However, from the coupling between the external magnetic field and the curvature of a surface, analytical solutions are predicted \cite{Vagson-PLA-376}.

In this paper, we show that the homogeneous DSG can be obtained for an arbitrary magnetically coated surface with rotational symmetry in the presence of an external magnetic field, provided the field is tuned with the curvature of the surface. The obtained solution consists in a $2\pi$ soliton, which induces a deformation in the rotationally symmetric surfaces due the presence of the magnetic field. Furthermore, we apply the model to three specific surfaces: cylinder, catenoid and hyperboloid, in order to obtain the characteristic lengths of the $2\pi$ soliton in these particular cases. The coupling between the magnetic field and the geometry of condensed matter systems is a interesting result and we believe that these results may guide future works in the control and manipulation of the shape and physical properties of magnetoelastic coated surfaces, superfluid helium, nematic liquid crystals, graphene bubbles, and topological insulators, once the results obtained here show 
 that magnetoelastic surfaces may be deformed in a specific point by the application of a magnetic field varying from a maximum value, when $\rho=0$, and tending to 0 when $\rho\rightarrow\infty$. 

This paper is outlined as follows: In section \ref{Model}, we present the adopted model; the results and discussions are presented in Section \ref{results}. Section \ref{examples} brings the application of the adopted model on three specific surfaces and in Section \ref{conclusions}, we present the conclusions and prospects.

\section{The model}\label{Model}

The total energy for a deformable, magnetoelastically coupled manifold is given by $E=E_{\text{mag}}+E_{\text{el}}+E_{\text{m-el}}$, where $E_{\text{mag}}$, $E_{\text{el}}$ and $E_{\text{m-el}}$ are the magnetic, elastic and magnetoelastic contributions to the total energy, respectively. The magnetic contribution is composed by the exchange, magnetostatic, anisotropy and Zeeman terms. In this work, we will focus our attention mainly on the exchange and Zeeman contributions to describe the magnetic properties of curved surfaces with rotational symmetry in the presence of an external magnetic field. In this case, the energy is well represented by the non-linear $\sigma$-model (NL$\sigma$M) on a surface in an external magnetic field:
\begin{equation}\label{GenHam}
H=\iint\left(\nabla\mathbf{m}\right)^2dS-g\mu\iint\mathbf{m}\cdot\mathbf{B}dS,
\end{equation}
where $\mathbf{m}$ is the magnetization unit vector ($\mathbf{m}^2=1$), $dS$ is the surface element, $\mathbf{B}$ is the applied magnetic field, $\mu$ is the magnetic moment, and $g$ is the $g$ factor of the electrons in the magnetic material.

This model was previously considered for studying the properties of a circular cylinder surface in the presence of an constant axial magnetic field and the well known DSG \cite{Leung-PRB-26} was derived from the Euler-Lagrange equations \cite{Saxena-PRB-58}. In that case, the authors have obtained a $2\pi$ soliton like solution and have shown that a geometrical frustration appears, due to a second length scale introduced by the magnetic field. A surface deformation has been predicted at the sector where the spins point in the opposite direction to the magnetic field. On the other hand, unlike the cylinder case, whenever the above model has been considered on an arbitrary curved surface, it have yielded the non-homogeneous DSG, whose solution could be obtained only numerically \cite{Dandoloff-JPhysA}. Here, we show that the homogeneous DSG system appears for any rotationally symmetric surface, if the magnetic field is pointing in the $z$-axis direction and its strength is a function of $\rho$, that is tuned with the curvature of the substrate. That is, $\mathbf{B}\equiv B(\rho)\mathbf{z}$, where $\rho\equiv\rho(z)$ is the radial distance from one point on the surface to the $z$ axis.

To proceed with our analysis, it will be useful to rewrite Eq.(\ref{GenHam}) on a general geometry with metric tensor $g_{ij}$, and rotational symmetry, described in cylindrical-like coordinate system. Thus, we have that $g_{\rho\phi}=g_{\phi\rho}=0$ and the Hamiltonian (\ref{GenHam}) is rewritten as:
\begin{equation}\label{HamCurv}
H=\iint\sqrt{\frac{1}{g^{\rho\rho}g^{\phi\phi}}}\left[g^{\rho\rho}\left(\partial_\rho\Theta\right)^2+g^{\phi\phi}
\sin^2\Theta\left(\partial_\phi\Phi\right)^2+g\mu B(\rho)(1-\cos\Theta)\right]d\rho d\phi,
\end{equation}
where $g^{\rho\rho}$ and $g^{\phi\phi}$ are the contravariant metric elements. The magnetization unit vector is parametrized by $\mathbf{m}=(\sin\Theta\cos\Phi,\sin\Theta\sin\Phi,\cos\Theta)$ and the rotationally symmetric curved surfaces are parametrized in cylindrical coordinates system $(\rho,\phi,z)$. Besides, we have assumed cylindrical symmetry for the order parameter, that is, $\Theta(\rho,\phi)\equiv\Theta(\rho)$ and $\Phi(\rho,\phi)=\Phi(\phi)$. It is important to note that, once the magnetic field is in the $z$-axis direction, the minimum energy for the Zeeman interaction is obtained when $\Theta=0$ and the maximum, when $\Theta=\pi$. As it will be discussed after in this text, this fact forces the surface to deform (in order to minimize the total energy of the system)  at the sector of the $2\pi$ soliton where the spins point in the opposite direction to the magnetic field.

\section{Results and discussions}\label{results}

We are interested in studying only static solutions of Heisenberg spins on curved surfaces in an external magnetic field. In this case, the Hamiltonian and Lagrangian of the system coincide, that is, $H=L$, where $L\equiv L[\Theta,\Phi,\dot{\Theta},\dot{\Phi},t]$ is the Lagrangian of the CMS. Thus, in order to obtain the ground and excited states for this system, we must solve the Euler-Lagrange equations (ELE) for the Hamiltonian (\ref{HamCurv}). Then, ELE yields:
\begin{equation}\label{EulerTheta}
\sqrt{\frac{g^{\rho\rho}}{g^{\phi\phi}}}\partial_\rho\left(\sqrt{\frac{g^{\rho\rho}}{g^{\phi\phi}}}\partial_\rho\Theta\right)=\frac{\left(\partial_\phi\Phi\right)^2}{2}\sin2\Theta+\frac{1}{g^{\phi\phi}}B'(\rho)\sin\Theta
\end{equation}
and
\begin{equation}\label{EulerPhi}
\sin^2\Theta\partial_\phi\left[\sqrt{\frac{g^{\phi\phi}}{g^{\rho\rho}}}\partial_\phi\Phi\right]=0,
\end{equation}
where $B'(\rho)\equiv g\mu B(\rho)$.

Since we are considering surfaces with rotational symmetry, the parametric equations associated with Eq.(3) and Eq.(4)  can be written, in cylindrical-like coordinates, such as $\mathbf{r}=(\rho\cos\phi,\rho\sin\phi,z(\rho))$,
where $\rho$ is the radius of the surface at height $z$, and $\phi$ accounts for the azimuthal angle. In this case, the covariant metric elements are given by:
\begin{equation}
g_{\phi\phi}=\frac{1}{g^{\phi\phi}}=\rho^2\hspace{1cm}\text{and}\hspace{1cm} g_{\rho\rho}=\frac{1}{g^{\rho\rho}}=z\,'^2+1,
\end{equation}
where $z\,'={\partial z}/{\partial \rho}$. From the above equations, it is easy to see that  $\partial_\phi\left(\sqrt{\frac{g^{\phi\phi}}{g^{\rho\rho}}}\right)=0$, and Eq.(\ref{EulerPhi}) is simplified to:
\begin{equation}\label{phieq}
\sin^2\Theta\partial^2_\phi\Phi=0.
\end{equation}

One can note that the above equation admits two kinds of solution: the first one is given by $\Theta=n\pi$, with $(n=0,\,1,\,2,\,...)$. In this case, if $n$ is even, the Zeeman term in the Hamiltonian (\ref{GenHam2}) is minimized, however, $n$ odd yields a maximization of the magnetic field interaction energy, in such a way that these solutions must be unstable, since the spins would be pointing in the opposite direction to the magnetic field. Thus, the solution $\Theta=(2n)\pi$ is the ground state of Eq. (\ref{GenHam}). However, we are interested in the existence of excited states, so called $2\pi$ solitons, which are a continuous transition between the two vacua, $\Theta=0$ and $\Theta=2\pi$, of this system.  This class of solutions have been already obtained for the cylinder surface \cite{Saxena-PRB-58}, however, analytical solutions have been also found  for other geometries, provided the magnetic field is coupled with the curvature of the substrate \cite{Vagson-PLA-376}. Here, the simplest way to obtain topological $2\pi$ soliton-like solutions is by considering the second simplest possible solution for the Eq. (\ref{phieq}), that is:
\begin{equation}
\Phi(\phi)=\phi+\phi_{_0}
\end{equation}
where $\phi_{_0}$ is a constant of integration that does not influence the energy calculations. Then, we can rewrite the Eq.(\ref{EulerTheta}) as:
\begin{equation}\label{GenThetaEuler}
\partial_\xi^2\Theta=\frac{\sin2\Theta}{2}+g_{\phi\phi}B'(\rho)\sin\Theta,
\end{equation}
where $d\xi=\sqrt{g^{\phi\phi}/g^{\rho\rho}}d\rho.$

From the analisys of  Eq.(\ref{GenThetaEuler}), there are two situations to be considered: firstly, when we take $B(\rho)=0$, it yields the single sine-Gordon equation and a topological soliton, which satisfies the self-dual equations, is obtained \cite{cylinder}. This particular value for $B(\rho)$ leads to the isotropic Heisenberg Hamiltonian, which has been previously studied on several curved geometries, e.g., torus \cite{torus}, cylinder \cite{cylinder,cylinder-2}, sphere \cite{sphere}, pseudosphere \cite{pseudosphere}, cone \cite{cone}, catenoid, hyperboloid \cite{Dandoloff-JPhysA}. It is important to note that, usually, the soliton presents a characteristic length scale (CLS) that depends on the geometrical properties of the underlying manifold. In general, the soliton's  CLS appears in front of the $\sin(2\Theta)$ term, in Eq. (\ref{GenThetaEuler}). However, this CLS is given by ($\sqrt{g^{\phi\phi}/g^{\rho\rho}}$), which is embedded in the $\xi$ para
 meter. Thus, the soliton would have its CLS rescaled to one, e.g., if we take the cylinder case, the CLS of the soliton is equal to the constant radius $\rho_{_0}$ of this surface. However, when we are working on an infinite cylinder, the change of variable $z\rightarrow z/\rho_{_0}$ eliminates any dependence on $\rho_{_0}$ \cite{cylinder}.

The second case to be discussed is given when we consider an arbitrary magnetic field $B(\rho)\neq0$. In this case, if the magnetic field is constant, and an arbitrary curved surface with rotational symmetry is considered,  Eq.(\ref{GenThetaEuler}) leads to the non-homogeneous DSG and numerical solutions are demanded \cite{Dandoloff-JPhysA}. However, it can be noted that if the external field is coupled with the surface curvature, in the form $B'(\rho)=g^{\phi\phi}B'_0$, where $B'_0=g\mu B(\rho_{_0})$ and $\rho_{_0}$ is the surface radius at $z=0$ plane, Eq.(\ref{GenThetaEuler}) yields the homogeneous DSG:
\begin{equation}\label{HomSGEq}
\partial_\xi^2\Theta=\frac{\sin2\Theta}{2}+B'_0\sin\Theta.
\end{equation}
Assuming that $\Theta(-\infty)=0$ and $\Theta(+\infty)=2\pi$, the solution for the above eqquation can be given by:
\begin{equation}\label{solution}
\Theta(\xi)=2\tan^{-1}\left(\frac{\rho_{_B}}{\zeta\sinh\frac{\xi}{\zeta}}\right),
\end{equation}
where $\rho_{_B}^2=1/B'_0$ and $\zeta=\rho_{_B}/(1+\rho_{_B}^2)^{1/2}$. Then, one can conclude that the homogeneous DSG must be obtained for the Heisenberg spins on curved surfaces in an external magnetic field only if the magnetic field is tuned to the surface curvature, varying in function of $1/\rho(z)$. 

Eq. (\ref{solution}) represents a $2\pi$ soliton, which is a topological excitation belonging to the second class of the second homotopy group, whose CLS is given by $\zeta$. Note that the increasing of $B'_0$ leads to the decreasing of the CLS of the soliton causing it to remain confined in smaller regions of the surface. It is also easy to see that $\zeta\rightarrow0$ when $B'_0\rightarrow\infty$. Note also that, if we put $B'_0=0$ in Eq.(\ref{solution}), a $\pi$ soliton is not obtained, as it happens when we consider Eq.(\ref{HomSGEq}) with $B'_0=0$. 
This is explained because these two solutions belong to two different homotopy classes and cannot be transformed one to another by continuous transformation or by limiting process  $B'_0\rightarrow 0$. It can also be noted that, by introducing a new CLS in the system, the magnetic field induces a geometrical frustration, and the soliton, which had its CLS rescaled to one, must choose a new CLS, given by $\zeta$, that is smaller than the length $\rho_{_B}$, introduced by the magnetic field.

In order to calculate the energy of the $2\pi$ soliton, we can rewrite the Hamiltonian (\ref{HamCurv}) as: 
\begin{equation}\label{GenHam2}
H'=H_1+H_2=2\pi\left[\int_{-\infty}^{\infty}\left[(\partial_\xi\Theta)^2+\sin^2\Theta\right]d\xi+
\int_{-\infty}^{\infty}\frac{1}{\rho_{_B}^2}(1-\cos\Theta)\right]d\xi,
\end{equation}
where $H_1$ is the part of the Hamiltonian that does not have dependence on the magnetic field and $H_2$ is the Zeeman term. In this case, the energy of the soliton is evaluated to give:
\begin{equation}\label{2piSolEn}
E_{_\text{S}}=8\pi\left[\left(1+\frac{1}{\rho_{_B}^2}\right)^{1/2}+\frac{1}{\rho_{_B}^2}\sinh^{-1}\rho_{_B}\right],
\end{equation}
that is larger than the minimum energy for the homotopy class with winding number $Q=2$, that is to say, $E_{_\text{2$\pi$ S}}=8\pi$. Only in the limit $\rho_{_B}\rightarrow\infty$, we get $E_{_\text{S}}\rightarrow8\pi$ and $\zeta\rightarrow1$. Once $\rho_{_B}=\sqrt{1/B'(\rho_{_0})}$, the energy of the soliton is associated with the strength of the magnetic field at the $z=0$ plane. In order to release the geometrical frustration introduced by the magnetic field, an elastic surface will deform to decrease the radius in the region where the soliton is centered. This will minimize the energy of the two terms that do not  depend on the magnetic field in the Hamiltonian (\ref{GenHam2}). Furthermore, by considering only the first term the development of ELE leads to:
\begin{equation}
\partial_\xi^2\Theta=\frac{\sin2\Theta}{2},
\end{equation}
as expected. This is the single sine-Gordon equation, whose solution is:
\begin{equation}
\Theta=2\arctan\left(\text{e}^\xi\right),
\end{equation} 
which represents a $\pi$ soliton, interpolating the two minima, $\Theta=0$ and $\Theta=\pi$, and which belongs to the first class of the second homotopy group. As it was mentioned before, the $\pi$ soliton solution can not be obtained if we take the limit $B\rightarrow0$ in the solution (\ref{solution}), as well as we do not obtain the $\pi$ soliton energy, predicted by the Bogomolnyi inequality \cite{Bogomol} by taking $B=0$ in the Eq. (\ref{2piSolEn}).

Now, we will focus on the last term in the Hamiltonian $H'$, which involve the external magnetic field:
\begin{equation}
H_2=2\pi\int_{-\infty}^{\infty}\frac{1}{\rho_{_B}^2}(1-\cos\Theta) d\xi.
\end{equation}
For the solution (\ref{solution}), $H_2$ will take the form:
\begin{equation}
H_2=4\pi\int_{-\infty}^{\infty}\frac{1}{\rho_{_B}^2}\left[\frac{1}{1+\left(1-\zeta^2\right)\sinh^2\frac{\xi}{\zeta}}\right]d\xi=4\pi\int_{-\infty}^{\infty}f(\rho_{_B},\xi)d\xi.
\end{equation}
This part of the Hamiltonian will decrease in energy if we locally decrease $\xi$, but keep it constant in Eq.(\ref{solution}). Therefore, for a magnetic field tuned with the surface, if we keep the surface cross section at $\xi\pm\infty$, the surface will deform in the region of the soliton. In the case of small magnetic field, we get:
\begin{equation}
E_{_\text{S}}=8\pi\left[1+\frac{1}{2\rho_{_B}^2}\left(1+2\ln{2\rho_{_B}}\right)\right]
\end{equation}
and for large magnetic field, we get $E_{_\text{S}}=8\pi/\rho_{_B}$.

As expected, the highest contribution for the magnetic energy density due to the interaction with the external magnetic field comes from the sector where the spins point in the opposite direction to the field. In this way, the $2\pi$ soliton would like to collapse and eliminate the region where the spins are opposite to the magnetic field. However, to align the spins, the system would fall in a $\pi$ soliton sector of the second homotopy group, which has infinite energy due to the interaction of the spins at $+\infty$ with the magnetic field. Thus, there is a hard-core repulsion between the two $\pi$ solitons in the system by virtue of the curvature. Due to the introduction, by the magnetic field, of a new CLS into the system, a geometric frustration appears in the problem and the energy in Eq.(\ref{2piSolEn}) is grater than $8\pi$. Then, the surface will try to deform in order to decrease its magnetic energy until it reaches the second homotopy class \cite{Saxena-PRB-58}. Obviously this deformation will cost elastic energy to the system, so the real deformation will balance between the gain in magnetic energy and the loss in elastic energy.This result suggests that we could, at first sight, deform an elastic membrane, coated with magnetic material, by using a variable magnetic field, whose strength would decrease with the increasing of the surface's rotational radius, $\rho$.

In the absence of an external field ($B_{_0}=0$), the self dual equation for the Eq.(\ref{HomSGEq}) is $\partial_\xi\Theta=\pm\sin\Theta$. For the DSG, we can write:
\begin{equation}\label{SDE}
\partial_{\xi}\Theta=\pm\sin\Theta\left[1+\frac{4\sin^2(\Theta/2)}{\rho_{_B}^2\sin^2\Theta}\right].
\end{equation}
The function in square brackets is greater than one for any value of $\Theta(\xi)$ and the self dual equation is not satisfied. Therefore, in order to satisfy the self dual equation and to lower the energy, the surface would deform. Note that all solutions of the Eq.(\ref{SDE}) satisfy the Euler-Lagrange equation, but not vice versa. The $2\pi$ soliton lattice of the double sine-Gordon equation in two regimes, $\rho_{_B}\leq1$ and $\rho_{_B}>1$, are similar to that found on the cylinder \cite{Saxena-PRB-58}.

\section{Some specific cases}\label{examples}

\begin{figure}\label{shapes}
\includegraphics[scale=0.35]{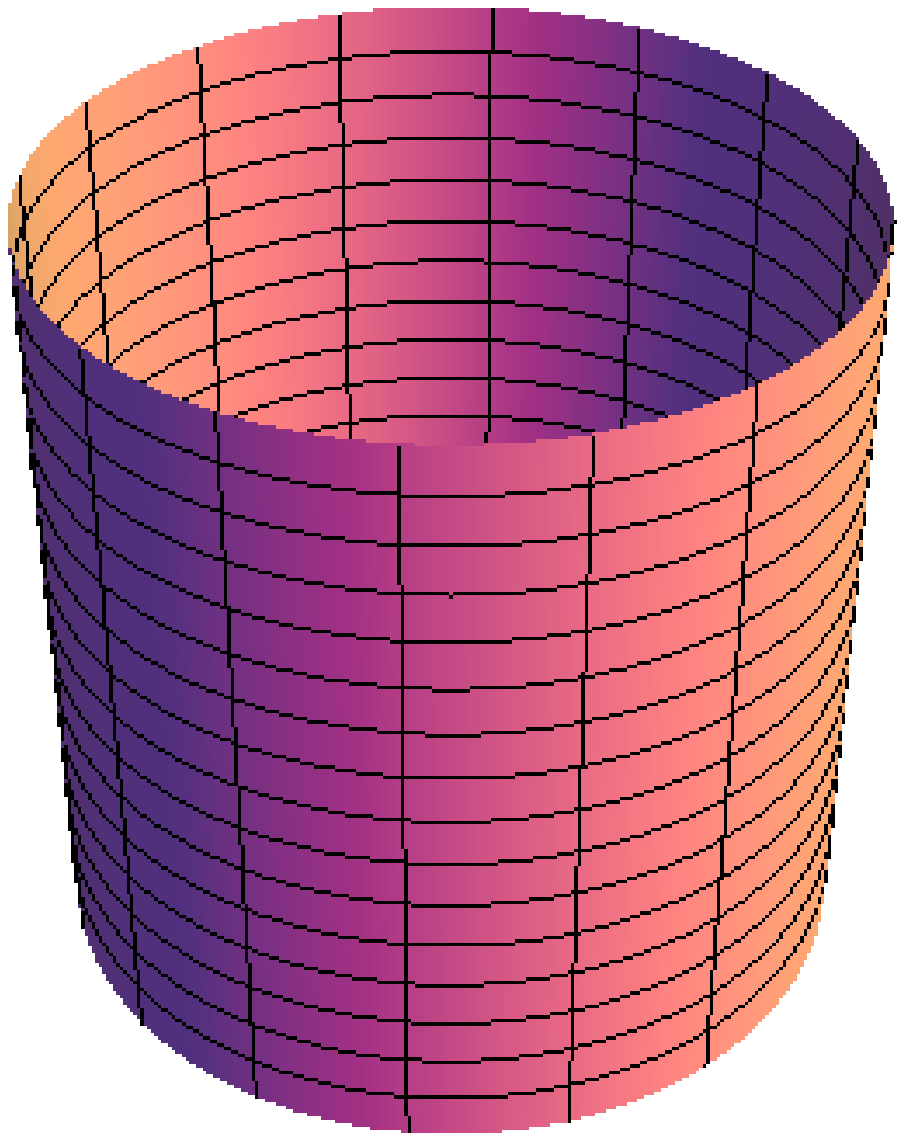}\includegraphics[scale=0.4]{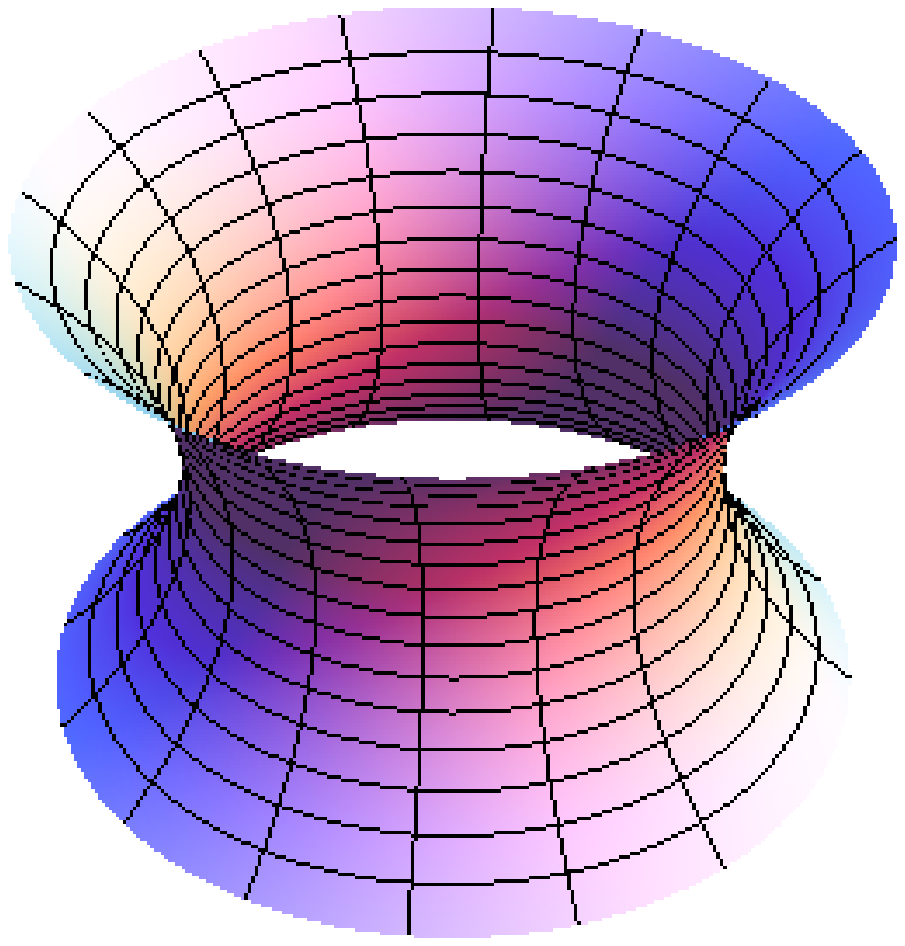}\includegraphics[scale=0.4]{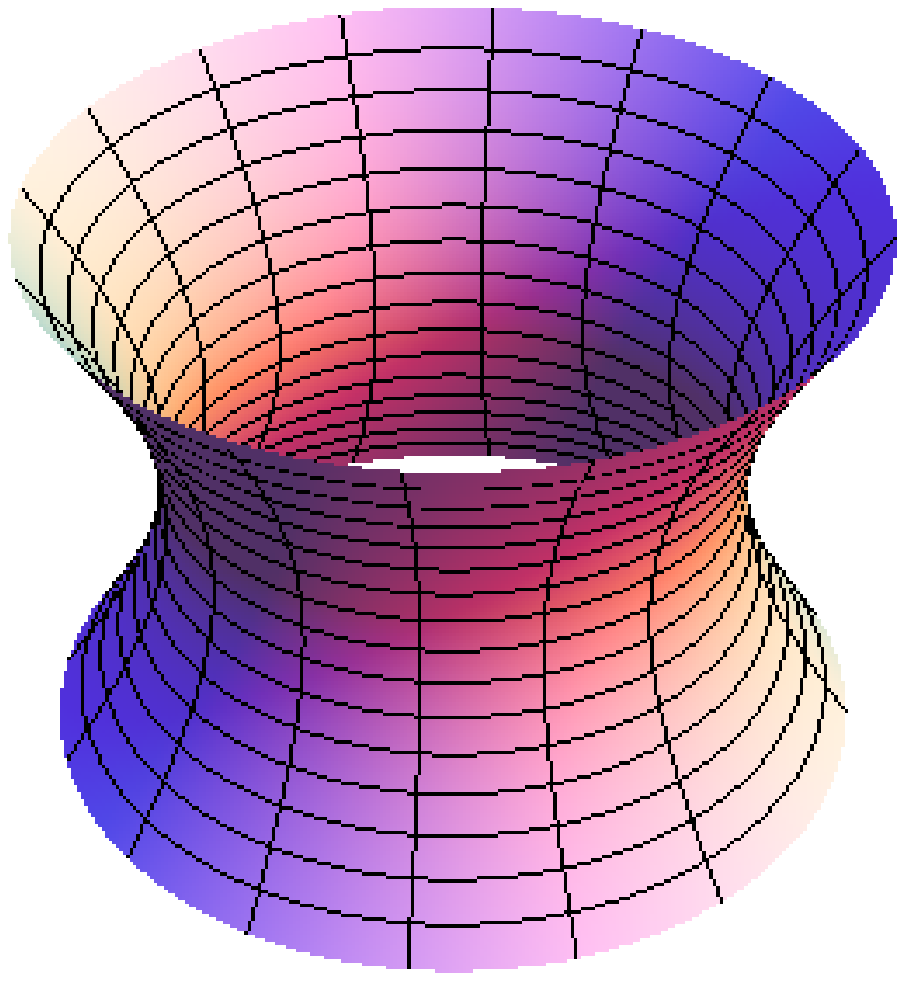}\caption{[Color online] From left to right, one can see the surfaces of cylinder, catenoid and hyperboloid. These are curved surfaces with rotational symmetry. In order to obtain a 2$\pi$-soliton on these surfaces, the magnetic field must be coupled with the curvature. In the case of the cylinder, the magnetic field is constant, while that one for the catenoid and hyperboloid varies with $\rho$, having maximum value in the $z=0$ plane, and vanishing its strength in $\rho\rightarrow\pm\infty$.}
\end{figure}

$2\pi$-soliton solutions should be found on infinite surfaces with rotational symmetry and the CLS is associated with the geometry of the substrate. In this case, we will apply our results on some infinite surfaces with this kind of symmetry in order to find the soliton CLS associated to each one of them. Here, we will consider three particular geometries: the cylinder, the catenoid and the hyperboloid, whose shapes are shown in Fig. \ref{shapes}. It can be noted that in spite of the shape similarity between the catenoid and hyperboloid surfaces, their geometric properties are different, as it will be discussed soon after. It is also important to note that the 2$\pi$-soliton, given by the solution (\ref{solution}) does not appear on finite surfaces, e.g., sphere, torus and/or barrels. In these cases, fractional solitons must be found, because the spin sphere will not be completely covered two times, as it should be in order to get an excitation belonging to the second class of the second homotopy group. In this way, topological arguments can not be used to ensure the stability of this excitation.

Thus, the first case to be analysed is the surface of the cylinder, which was previously studied by Saxena \textit{et al} \cite{Saxena-PRB-58}. In that work, the authors have predicted the appearing of a 2$\pi$-soliton on the cylinder in the presence of a constant magnetic field and have found its CLS as a function of $z$. The cylinder presents null Gaussian curvature  and its mean curvature is given by $M=1/2r$. In this case, we have that $z(\rho)\equiv z$ is a constant, thus, the cylindrical coordinates parametrization leads to the metric elements:
\begin{equation}
g_{\phi\phi}^\text{cyl}=\rho^2=r^2\hspace{1cm} \text{and} \hspace{1cm}g_{\rho\rho}^\text{cyl}=1,
\end{equation}
where $r$ is the radius of the cylinder. It is easy to see that the magnetic field to be applied on the surface is $B'(\rho)_{_\text{cyl}}=({1}/{r^2})B'_0$, which is constant. The CLS is easily calculated to be:
\begin{equation}
\xi_{_\text{cyl}}=\ln{r}.
\end{equation}
As expected, the result here obtained would be the same as that given in Eq.(4) of Ref. \cite{Saxena-PRB-58} if we had considered $z$ instead $\rho$ as the variable of the problem. 

From the analysis of Eq. (\ref{GenThetaEuler}), we could conclude that the homogeneous DSG can be obtained from the Heisenberg model in an external field for any other rotationally symmetric surface, besides the cylinder, if the magnetic field has its strength varying with the metric element $g_{\phi\phi}$. On the other hand, if we apply a constant magnetic field on an arbitrary curved magnetoelastic surface, the solutions to the model described by the equation (\ref{GenHam}) lead to the inhomogeneous DSG, which must be solved numerically \cite{Dandoloff-JPhysA}. We continue our analysis, considering two surfaces with negative varying Gaussian curvature, namely the catenoid and the hyperboloid.

The catenoid is a minimal surface, with its mean curvature equals zero everywhere. However, unlike the cylinder surface, its Gaussian curvature does not vanish. Indeed, the the catenoid presents a negative Gaussian curvature, which is given by:
\begin{equation}
K=-\frac{1}{\rho^2}\text{sech}^4\left(\frac{z}{\rho}\right).
\end{equation}
Another difference of the catenoid in relation to the cylinder is the act that the radius of the catenoid is not constant, but it varies as a  function of $z$, as follows:
\begin{equation}\label{rhocat}
\rho(z)=r\cosh(z/r).
\end{equation} 
The catenoid can be parametrized by $(\rho\cos\phi,\rho\sin\phi,r\cosh^{-1}(\rho/r))$, where $r$ is the radius of the surface at the $z=0$ plane. This parametrization leads to:
\begin{equation}
g_{\phi\phi}^\text{cat}=\rho^2\hspace{1cm}\text{and}\hspace{1cm} g_{\rho\rho}^\text{cat}=\frac{\rho^2}{\rho^2-r^2}.
\end{equation}
Then, in order to obtain the homogeneous DSG, the magnetic field to be applied on this surface,  will be $B'(\rho)_{_\text{cat}}=({1}/{\rho^{2}})B'_0$. Once $\rho$ is given by Eq. (\ref{rhocat}), the magnetic field is not constant, but it varies in function of $z$. Indeed, for $\rho\rightarrow\infty$, the field tends to zero and its maximum value ($B'(\rho=r)_{_\text{cat}}=({1}/{r^2})B'_0$) must occur at the point $\rho(0)=r$. Finally, it is easy to calculate the CLS on the catenoid in an external varying magnetic field, which admits a 2$\pi$-soliton solution with: 
\begin{equation}
\xi_{_\text{cat}}=\ln\left[2\left(\rho+\sqrt{\rho^2-r^2}\right)\right].
\end{equation}

The last surface on which our model will be applied is the hyperboloid, which has a shape similar to the catenoid, however, while the catenoid has mean curvature zero everywhere, the hyperboloid has variable mean and Gaussian curvatures \cite{Math-Wolfram}.
A one-sheeted hyperboloid can be parametrized by $(\rho\cos\phi,\rho\sin\phi,({b}/{r})\sqrt{\rho^2-r^2})$, where, again, $r$ is the radius of the surface at $z=0$ and $b$ is a multiplicative parameter that accounts for the height of the surface. This parametrization leads to: 
\begin{equation}
g_{\phi\phi}^{\text{hyp}}=\rho^2\hspace{1cm}\text{and}\hspace{1cm} g_{\rho\rho}^{\text{hyp}}=\frac{\rho^2\left(b^2+r^2\right)-r^4}{r^2\left(\rho^2-r^2\right)}.
\end{equation}
From now on, in order to obtain analytical solutions, we will use the biharmonic coordinate system (BC), in which $b=r$, to describe a particular kind of hyperboloid \cite{PolarHyp}, here called polar hyperboloid. In this case, the metric element simplifies to: 

 \begin{equation}
g_{\rho\rho}^{\text{phyp}}=\frac{2\rho^2-r^2}{\rho^2-r^2}
\end{equation}

and, to obtain the homogeneous DSG, the magnetic field must be given by $B'(\rho)_{_\text{hyp}}=({1}/{\rho^2})B'_0$. Thus, as well as in the catenoid case, in order to obtain a 2$\pi$-soliton, the  magnetic field must be $\rho$ dependent, having its maximum value, given by $B'(\rho=r)_{_\text{hyp}}=({1}/{r^2})B'_0$, at the $z=0$ plane. It is easy to show that the magnetic field applied on the hyperboloid described by BC varies with $\rho=\sqrt{r^2+z^2}$, then, $B'(\rho)_{_\text{hyp}}\rightarrow0$ when $z(\rho)\rightarrow\pm\infty$. Finally, the polar hyperboloid also admits the solution given by the equation (\ref{solution}), however, the $\xi$ parameter for this surface is given by a large and tedious expression which we will not consider here. Despite the fact that $\lim_{\rho\rightarrow\infty}B(\rho)=0$, it is important to note that the total magnetic flux through these infinite surfaces diverges as $\ln\rho$ at infinity for the catenoid and hyperboloid, but remains constant and depending on $r$ for the cylinder.

\section{Conclusions and prospects}\label{conclusions}

In conclusion, we have shown that the Euler-Lagrange equations derived from the continuum approach for classical Heisenberg spins on a surface with rotational symmetry in an external magnetic field, pointing in the $z$ axis direction, is the homogeneous DSG provided the field is coupled with the curvature of the surface. We have found a single $2\pi$-soliton like solution for this model for an arbitrary surface. Surface deformations at the sector where the spins point in the opposite direction to the magnetic field were predicted, in order to diminish the energy, which is greater than that predicted by the Bogomolnyi's inequality. 

We have applied this model on three specific surfaces: cylinder, catenoid and hyperboloid, which have different geometrical properties, characteristic length scales and Gaussian curvatures. As expected, each considered surface admits a 2$\pi$-soliton, but the magnetic field which gives the homogeneous DSG presents different behavior on each one (in each case B is a function of $\rho(z)$). While the field to be applied on the cylinder must be constant, the field  which leads to the homogeneous DSG on the catenoid and the hyperboloid is inversely proportional to $\rho(z)$. Since the characteristic length of the soliton depends on the strength of the field, the magnitude of the deformation can be controlled by the magnitude of the applied magnetic field. Finite surfaces admit fractional 2$\pi$-solitons, which are not topologically stable.

Our results can play an important role for experimental studies where magnetoelastic effects here predicted could be observed. For example, a membrane coated by a magnetic material could be deformed by using this technique. Additional theoretical work which also  includes surface tension could clarify the deformation dynamics of magnetic coated membranes and/or magnetoelastic metamaterials \cite{Lapine-Nature}. Finally, we believe that this issue may be relevant for future studies on manipulation and control of the shape and physical properties of nematic liquid crystals, curved graphene sheets and topological insulators.

\section*{Acknowledgements}
We thank the Brazilian agency CNPq (grant number 562867/2010-4) and Propes of the IF Baiano, for financial support. Carvalho-Santos thanks J. D. Lima, P. G. Lima-Santos and G. H. Lima-Santos for encouraging the development of this work.

\end{document}